\documentclass[twocolumn]{revtex4}
\pdfoutput=1
\usepackage{graphicx}
\usepackage{ulem}
\usepackage{color}
\newsavebox{\mybox}
\newcommand\origcite{}
\let\origcite\cite
\renewcommand\cite[1]{\sbox{\mybox}{\origcite{#1}}\usebox{\mybox}}
\definecolor{orange}{rgb}{1.0,0.5,0.0}
\newcommand\jydAdd[1]{#1}
\newcommand\jydDel[1]{}

\begin{document}

\title{Coexisting multiple dynamic states generated by magnetic field
in Bi$_2$Sr$_2$CaCu$_2$0$_{8+\delta}$ stacked Josephson junctions}

\author{Yong-Duk~Jin}
\affiliation{Department of Physics, Pohang University of Science
and Technology, Pohang 790-784, Korea}
\author{Hu-Jong~Lee}
\email[Corresponding author: ]{hjlee@postech.ac.kr}
\affiliation{Department of Physics, Pohang University of Science
and Technology, Pohang 790-784, Korea}
\author{A.~E.~Koshelev}
\affiliation{Materials Science Division, Argonne National
Laboratory, Argonne, Illinois 60439, USA}
\author{Gil-Ho~Lee}
\affiliation{Department of Physics, Pohang University of Science
and Technology, Pohang 790-784, Korea}
\author{Myung-Ho~Bae}
\affiliation{Department of Electrical and Computer
Engineering, Micro and Natotechnology Laboratory, University of
Illinois at Urbana-Champaign, Illinois 61801, USA}

\begin{abstract}
Josephson vortices in naturally stacked
Bi$_2$Sr$_2$CaCu$_2$O$_{8+\delta}$ tunneling junctions display
rich dynamic behavior that derives from the coexistence of three
basic states: static Josephson vortex lattice, coherently moving
lattice, and incoherent quasiparticle tunneling state. Rich
structure of hysteretic branches observed in the current-voltage
characteristics can be understood as combinatorial combinations of
these three states which are realized in different junctions and
evolve separately with magnetic field and bias current. In
particular, the multiple Josephson-vortex-flow branches at low
bias currents arise from the individual depinning of Josephson
vortex rows in each junction.
\end{abstract}


\maketitle

\jydAdd{Natural Josephson junctions (JJs), closely
packed in an atomic scale, form along the c axis of highly
anisotropic Bi$_2$Sr$_2$CaCuO$_{8+\delta}$ (Bi-2212)
superconductors \cite{KleinerPRL92a}. Josephson vortices (JVs) are
generated in these naturally stacked JJs in an in-plane magnetic
field.} Since the period between adjacent CuO$_2$ bilayers, $s$
(=1.5~nm), is much smaller than the in-plane component of the
London penetration depth $\lambda_{ab}$ \jydAdd{($\sim$ 200 nm)},
a Josephson vortex (JV) spreads over many junctions, which leads
to inductive coupling between JVs. In stacked JJs, JVs fill every
junction in a field higher than $B_{cr} = \Phi_0/(2\pi\gamma s^2)$
($\sim$0.75~T for Bi-2212), where $\gamma$ \jydAdd{($\equiv
\lambda_{c}$/$\lambda_{ab} \sim$250; $\lambda_{c}$ is the
out-of-plane penetration depth)} is the magnetic anisotropy ratio.
In this high-field region, JVs configurations in the static state
are well understood and are known to be in a triangular lattice
\cite{BulaevskiiPRB91a,XHuPRB04a,NonomuraPRB06a,KoshelevPRB06a}.
Despite much interest, however, dynamic state of JVs is far less
understood although various aspects of dynamical properties are
proposed including the possible lattice structures
\cite{KoshelevPRL00a,ArtemenkoPRB03a,MHBaePRB07a}, interaction
between JVs and electromagnetic
excitations\cite{KleinerPRB94b,PedersenPRB98a,MachidaPhysicaC00a,SMKimPRB05a},
and coherent characters of the JV motion \cite{LatyshevPRL01a}
over the whole junctions in a stack.

Key elements governing the dynamics of JVs are well reflected in
the tunneling current-voltage (I-V) characteristics, which reveal
the relation between the driving force acting on JVs in a junction
by the bias current and the responsive JVs motion that induces a
voltage drop across the junctions. For instance, JV viscosity is
determined by the in- and out-of-plane quasiparticle dissipation
that is extracted from the tunneling JV-flow resistance
\cite{KoshelevPRB00a,LatyshevPRB03a}. Oscillatory tunneling
magnetoresistance is also used to confirm the coherently moving JV
lattice and the influence of the edge barrier potential on the JV
motion \cite{OoiPRL02a,KoshelevPRB02a,MachidaPRL03a}. Interaction
between JVs and electromagnetic excitations is another interesting
subject of JV dynamics where the resonance with cavity modes
appears as Fiske steps \cite{KrasnovPRB99a,SMKimPRB05a}. Moreover,
resonance between collectively moving JVs and plasma mode
excitations in naturally stacked JJs has been studied extensively
\cite{KleinerPRB94b,PedersenPRB98a,MachidaPhysicaC00a} where
multiple subbranches in the I-V characteristics have been claimed
to be an experimental evidence for the
resonance\cite{MHBaePRB07a}. The interaction between Josephson and
pancake vortices (PVs) 
is also a high focus of recent studies
\cite{GrigorenkoNature01a,BulaevskiiPRB92a,KoshelevPRL99a,KoshelevPRB06a}.
It has been demonstrated that the motion of PVs can be manipulated
by the JVs via attractive interaction between them
\cite{ColeNatureMater06a,TokunagaPRB02a,VlaskoPRB02a}. This
suggests that the motion of JVs can alternatively be influenced by
the PVs\cite{KoshelevPRB06a}.

In this study we report that the JV-flow characteristics become a
lot more diverse when pinning effect is introduced. In an in-plane
magnetic field of $\sim$1~T on naturally stacked JJs, JVs are
pinned down for a low bias current but get depinned separately in
different junctions at a current higher than the depinning
current. The pinning and depinning of JVs give rise to
sub-branches in the I-V characteristics, which arise as a
combination of three distinct junction states: static JV lattice,
coherently moving lattice, and incoherent quasiparticle tunneling
(IQT) state. With increasing in-plane magnetic field strength over
2~T, the sub-branches become separated from IQT branches due to
reduction of the depinning current and form JV-flow branches
(JVFBs). \jydAdd{The coexisting multiple dynamic junction states
are} better resolved as the pinning of JVs by PVs is enhanced in a
slightly tilted in-plane magnetic field. \jydAdd{It allows an
unprecedented accurate access to JV dynamic states in the
atomically stacked tunneling junctions.} A comparison between the
JV-flow and the zero-field IQT curves indicates that the
coherently moving JVs form a triangular lattice configuration.

\section{Experiment}
We prepared slightly overdoped Bi-2212 single crystals by the
standard self-flux method \cite{NamKimPRB99a}. Three rectangular
stacks of junctions sandwiched between two gold-layer electrodes
were fabricated [see the insets of Fig. 1(c)] by using the
double-side cleaving technique \cite{WangAPL01a,MHBaeAPL03a} in
combination with the electron-beam lithography and the thermal
evaporation. The superconducting transition of the c-axis
tunneling conductance at $T_c$ (=87~K) indicates the oxygen doping
concentration \cite{AllgeierPhysicaC90a} to be $\sim$0.19. The
lateral size of the three samples was
$1.5\times10$~$\mu\textrm{m}^2$, which was in the long-junction
limit as the length was longer than the Josephson penetration
depth $\lambda_J$ ($=\gamma s \sim 0.3$ $\mu\textrm{m}$ for
Bi-2212). An in-plane external magnetic field was applied
perpendicular to the long side of a sample. We report on the
typical c-axis tunneling I-V characteristics from one of the
samples, which arose from the tunneling-current-driven JV-flow
dissipation. The total number of junctions $N$ was determined to
be 23 from the number of zero-field IQT branches \cite{QTBs} in
the inset of Fig.\ 1(a) (the suppressed two lowest-voltage
branches are from the surface junctions \jydAdd{with the reduced
Josephson coupling} \cite{NamKimPRB99a}). The contact resistance
($\sim$10~$\Omega$) included in our two-probe measurements can be
safely neglected as it is about two orders of magnitude smaller
than the tunneling resistance of each junction. The sample was
field cooled without a bias current whenever the field strength or
the field angle was altered. The in-plane field direction was
tuned within the accuracy of 0.01$^\circ$ by finding the angle for
maximum JV-flow resistance \jydAdd{at temperature around 65 K}
while controlling a stepper motor placed at room temperature.
\jydAdd{As the pinning on JVs by PVs is minimized at the
best-aligned field angle the JV-flow resistance becomes maximum.}

\section{Results and analysis}
Figure 1 shows the gradual variation of the I-V curves for $H$
varying from 1 to 3~T, where the fields were best aligned to the
in-plane direction. In a low magnetic field of 1~T [Fig.\ 1(a)]
the IQT branches remain conspicuous with critical switching
currents much reduced from the zero-field values shown in the
inset of Fig.\ 1(a) for all the branches. For this field value,
one also notices the development of tiny kink structure with some
indistinct subbranches for $I\sim10$~$\mu$A, which will be
discussed in detail below. For 1.2~T, subbranches develop around
the kink so that branching becomes very crowded [Fig.\ 1(b)]. With
further increasing field these subbranches move to a lower-bias
current region and get separated from the IQT branches for higher
bias currents [Figs.\ 1(c) and (d)]. It will be shown below that
the subbranches are directly linked to the JV motion and thereby
constitute the JVFBs. It has been suggested
\cite{KleinerPRB94b,PedersenPRB98a} and claimed to be confirmed
\cite{ThyssenIEEE97a,MHBaePRB07a} that the collectively moving JVs
in stacked JJs resonate with transverse Josephson plasma modes,
which generates the multiple JVFBs. But, with increasing the
in-plane magnetic field \jydAdd{the switching currents in JVFBs
keeps shrinking [see Fig.\ 1(d) and its inset]. The JV resonance
picture does not predict the field dependence of the switching
current so that this trend cannot be explained by the JV resonance
picture \cite{KleinerPRB94b}.}

\begin{figure}[htb]
\begin{center}
\leavevmode
\includegraphics[width=0.9\columnwidth]{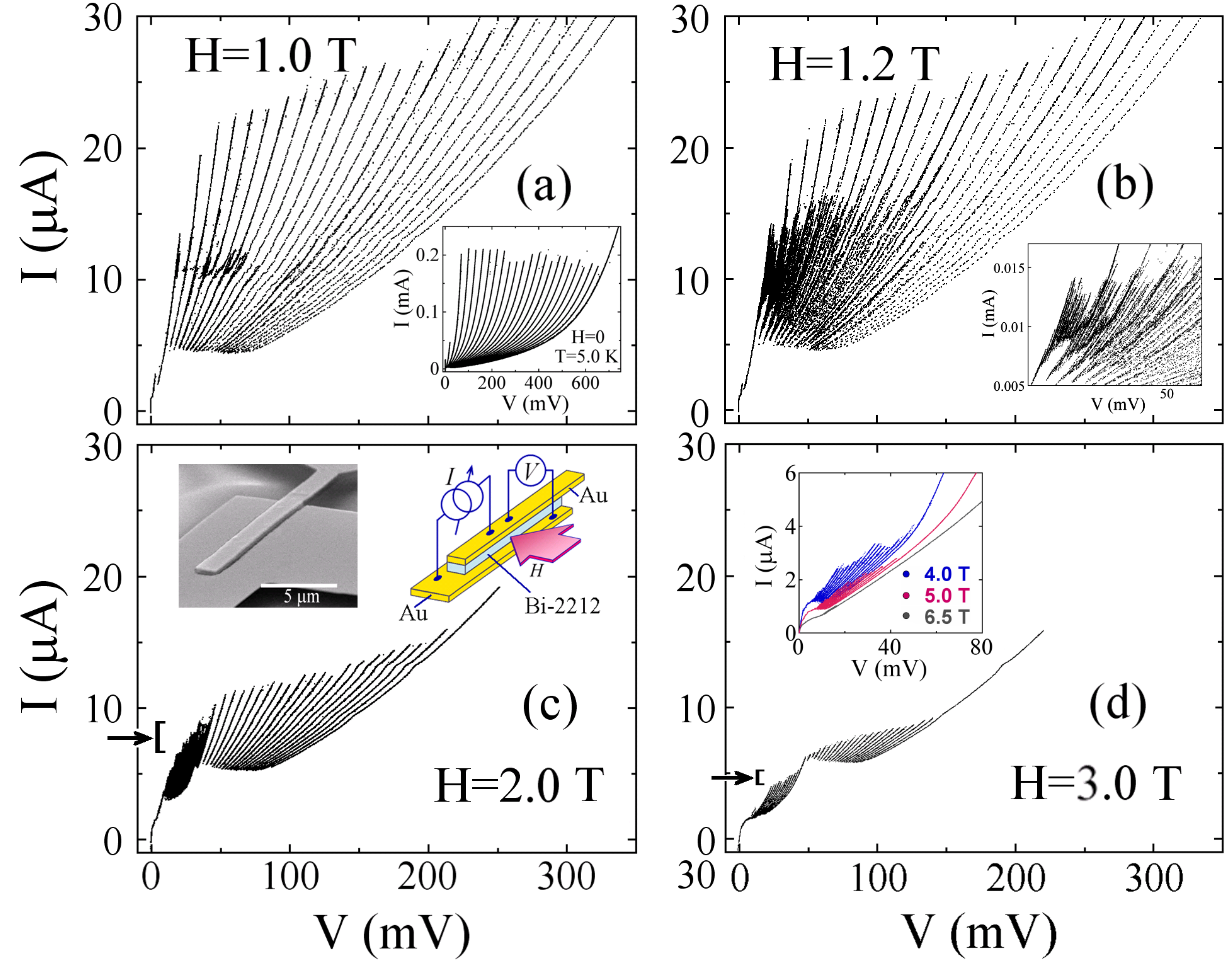}%
\caption{(colour online) (a)-(d) Magnetic-field dependence of the
Josephson-vortex-flow (JVF) I-V curves at 4.3 K for the
best-aligned planar fields of 1 to 3 T. Inset of (a): zero-field
quasiparticle branches in tunneling I-V curves. Inset of (b):
close-up feature of the low-bias subbranching for 1.2 T. Insets of
(c): (left) scanning electron microscopic picture of the sample
stack, (right) schematic measurement configuration. Inset of (d):
JVF branches for fields of 4, 5, and 6.5 T. The arrows in the
insets of (c) and (d) denote the depinning currents (see the text
for details).} \label{Fig:FieldSweep}
\end{center}
\end{figure}

Figure 2 illustrates the close-up details of Fig.\ 1(a) for 1~T
(red and green curves), together with the zero-field IQT curves
(gray ones). The red and green curves are bordered around
$I$$\sim$10~$\mu$A. In the low-bias red-curve region the 1~T
branches are in exact coincidence with the zero-field IQT
branches. This implies that JVs are completely pinned down without
showing any JV-flow voltages and thereby constitute the
\textit{static-vortex state} \cite{PinningOrigin}. This also
implies that the magnetic field very weakly influences the
resistive junctions. As one moves to more right branches,
junctions containing static JVs (thus, in the zero-voltage state)
turn into the IQT state one by one (see the lower schematic JV
configurations), causing the voltage jumps between neighboring
branches.

\begin{figure}[htb]
\begin{center}
\leavevmode
\includegraphics[width=0.9\columnwidth]{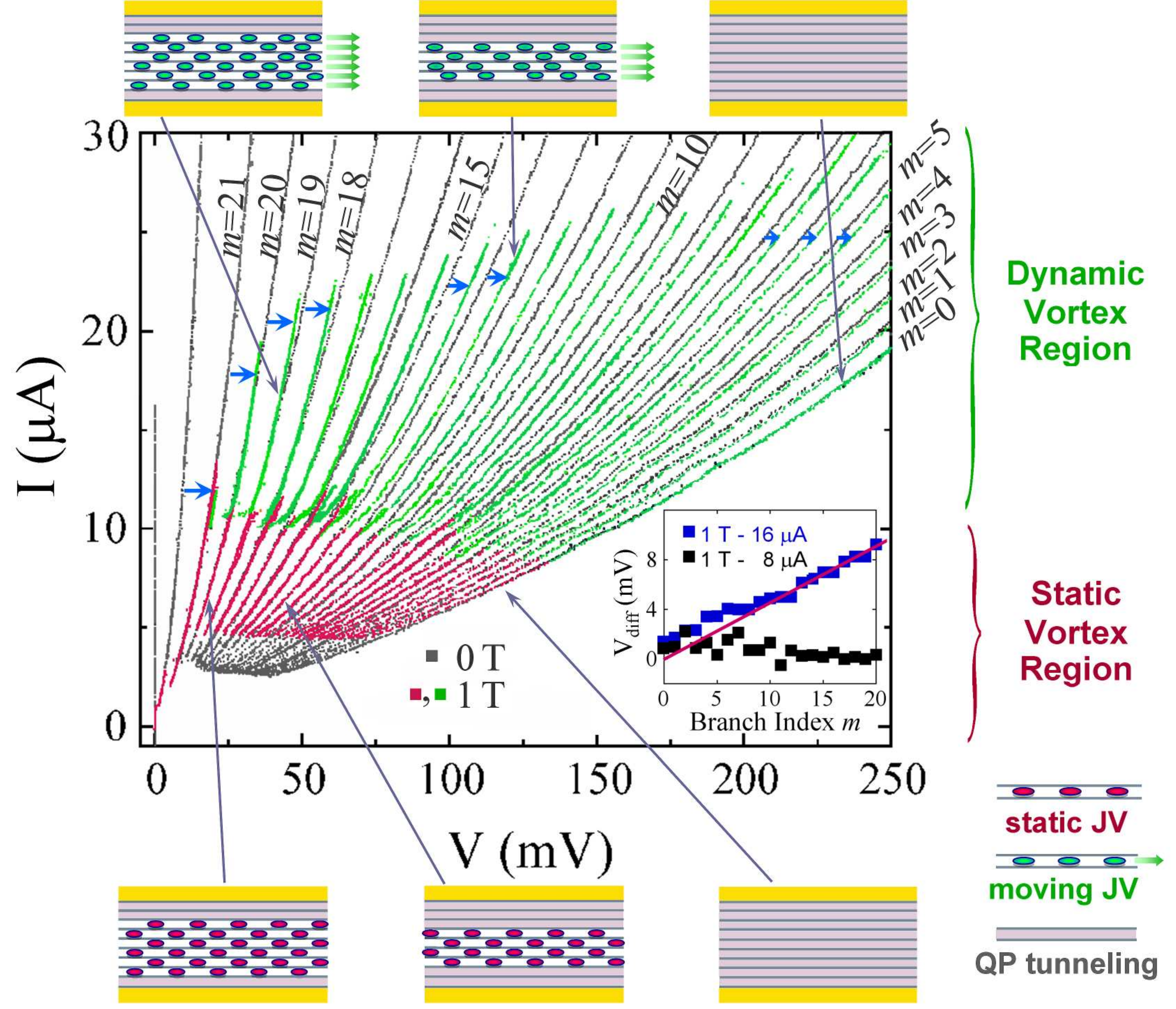}%
\caption{(colour) Detailed I-V curves in an in-plane field of 1~T
(red and green curves) along with the zero-field IQT (grey)
curves. Schematic vortex configurations corresponding to a few
branches are illustrated. Inset: the voltage difference
($V_{diff}$) between neighboring in-field and zero-field branches
for each $m$ for the given bias currents. The line is a guide to
the eyes.} \label{Fig:1TDetail}
\end{center}
\end{figure}

On the other hand, the green-curve region represents the
\textit{dynamic-vortex state}, where the JVs are depinned by a
higher bias current and thus each IQT branch is shifted by the
corresponding JV-flow voltage as denoted by blue arrows for some
branches. The inset of Fig.\ 2 shows the voltage difference
($V_{diff}$) between a zero-field IQT branch and the corresponding
branch in 1~T for each value of $m$ at the given bias currents.
The branch index $m$ denotes the number of junctions containing
dynamic vortices. For $I$=8~$\mu$A in the red-curve region,
$V_{diff}$ almost uniformly vanishes, supporting that the JVs in
all the junctions are static in this region. For $I$=16~$\mu$A in
1~T, on the other hand, $V_{diff}$ rises linearly with increasing
$m$, satisfying the relation $V_{diff}=V_m(H;I)-V_m(0;I)=m \delta
V(H,I)$. This indicates that, as more junctions turn into the
dynamic-vortex state, all the depinned JVs move with a uniform
velocity, \jydAdd{which is consistent with the previous studies on
the Shapiro step response in the coherent JV-flow state}
\cite{LatyshevPRL01a}. Here, $\delta V(H,I)$ (\textit{e.g.},
0.44~mV at 16~$\mu$A) is the average JV-flow voltage contribution
of each junction for a given field and a current.  Thus, the kink
near 10~$\mu$A in the 1~T I-V curves represents the boundary
between the regions of static JVs (showing pure IQT branches) and
the dynamic JVs (showing branches of IQT and moving JVs).

In Fig.\ 2, the static-JV state (red curves) for high $m$ values
abruptly turns into the corresponding dynamic-JV state (green
curves). But, in fact, the fine subbranches near the kink seen in
Figs.\ 1(a) and 2 signify the gradual transition between the two
JV states as JVs are depinned separately in each junction. Each
subbranch corresponds to different number of junctions containing
moving JVs. Subbranches become clearer for a slightly higher
magnetic field of 1.2~T as seen in the inset of Fig.\ 1(b). Here,
a higher-$m$ branch has more combinatorial configurations of
junctions with static and dynamic JVs. That causes a branch
located more left to split into a higher number of subbranches as
in the inset of Fig.\ 1(b). With increasing fields these
subbranches become separated from IQT branches, developing into
the JVFBs in higher magnetic field above $\sim$2~T [Figs.\ 1(c)
and (d)]. This feature can be qualitatively understood if one
assumes that the main pinning force acting on JVs, in this
relatively low in-plane field range, is from any field-independent
pinning sources\cite{PinningOrigin}. Then the total Lorentz force
on JVs in a bias current increases for a higher in-plane field as
more JVs are introduced to a junction while the pinning force
remains insensitive to the in-plane field strength. In
consequence, the subbranches move to a lower-bias current region
and get separated from the IQT branches when the depinning
currents [denoted by the arrows in Fig.\ 1(c) and (d)] become
smaller than the retrapping current of the JJs, at which the
junctions in the IQT state turn into the Cooper-pair tunneling
state.

Detailed development of the JVFBs is more clearly demonstrated in
the presence of pancake vortices (PVs). A PV in a superconducting
CuO$_2$ plane, which can be easily pinned down by defects in a
crystal such as the oxygen vacancies, tends to exert a pinning
force on adjacent JVs by the attractive interaction
\cite{KoshelevPRB06a,BulaevskiiPRB96a}. Figures 3(a)--(d)
illustrate the variation of the JV-flow I-V curves with slightly
tilting the sample stack with respect to the best-aligned in-plane
position in 4~T. The JVFBs extend with increasing the c-axis field
component $H_{(c)}$ along with the slight increase of the tilt
angle [$H_{(c)}$=350~Oe for $\theta$=0.5$^\circ$]. The shape of
the JVFBs was highly symmetric with respect to $\theta$=0$^\circ$
(not shown). For a higher tilt angle, larger bias currents are
required to depin JVs from the increased population of PVs, which
extend the JVFBs into a higher bias current region. This
PV-pinning effect on the JVFBs was also reproduced as the PVs were
generated in a low c-axis-oriented field (an order of a few tens
Oe) by a separate Helmholtz coil with the main solenoid at the
best-aligned in-plane field position (not shown).

\begin{figure}[htb]
\begin{center}
\leavevmode
\includegraphics[width=0.9\columnwidth]{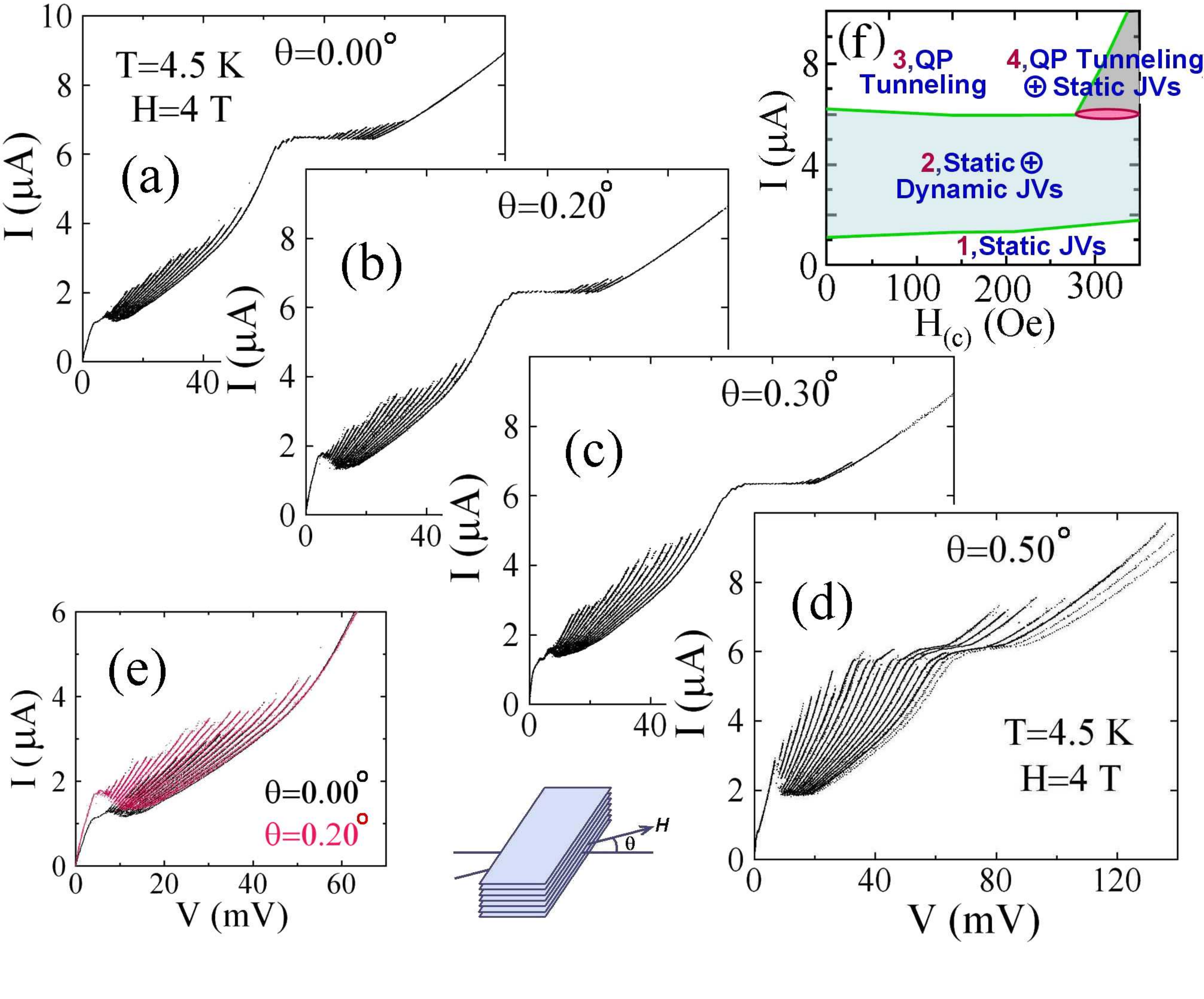}%
\caption{(colour online) (a)--(d) Progressive variation of the
Josephson-vortex(JV)-flow I-V curves measured at 4.5~K as the
sample stack is slightly tilted up to $\theta$=0.5$^\circ$ from
the in-plane field position in 4~T. (e) A comparison of JV-flow
I-V curves for $H$=4~T at the best-aligned position and at
$\theta$=0.2$^\circ$. (f) Different JV regimes for the near-planar
field of 4 T, as a function of c-axis field component of up to
350~Oe (or $\theta$=0.5$^\circ$).} \label{Fig:AngleSweep}
\end{center}
\end{figure}

Once a JV is depinned, the PVs exert a far less effect on the JV
motion for low tilt angles although the presence of higher-density
PVs in a higher tilt angle may slow down the JV motion
\cite{KoshelevPRB06a}. This bears an analogy with the fact that
the dynamic friction coefficient is much smaller than the static
one. Thus, JVFBs for two different field angles almost overlap as
seen in Fig.\ 3(e). In this situation, as the depinning current
increases for a higher tilt angle, the maximum voltage at the
switching current of each JVFB is also enhanced. But, this feature
is in contradiction to the picture of the resonance of JVs with
the transverse plasma modes \cite{KleinerPRB94b} as the cause of
branching of JV-flow I-V curves. In this picture, the switching
voltage between branches corresponds to the resonant propagation
mode velocity of the transverse Josephson plasma excitation.
\jydAdd{For a fixed value of an in-plane field with a constant JV
number density, the switching voltage should be insensitive to
$H_{(c)}$ as the resonant mode velocity, which is proportional to
$\omega_p \times \lambda_J$ ($\omega_p$; Josephson plasma angular
frequency) \cite{MachidaPhysicaC00a}, is independent of the
Josephson critical current $I_c$.}

As $\theta$ exceeds 0.4$^\circ$ [Fig.\ 3(d)], the branch structure
is modified considerably so that JVFBs become much more pronounced
and extend into the ``IQT-branch region'', which will be referred
to as \textit{``extended-JV branches''}. As it will be discussed
in detail below, this extended-JV branches arise from the combined
distribution of junctions with static JVs and the ones in the IQT
state.

Figure 4 illustrates JV configurations of Fig.\ 3(d). The
lowest-voltage single branch (for $I$$<\sim$3~$\mu$A) represents
the state where all the stacked junctions, except for the two
surface junctions, retain only static JVs [configuration (i)]. The
finite voltage of the branch in this state is generated by the IQT
in the two surface junctions. This low-bias single branch, with a
static triangular JV configuration, however, showed
magnetoresistance oscillations with the periodicity of a
half-flux-quantum entry per junction
\cite{OoiPRL02a,KoshelevPRB02a,MachidaPRL03a}, when JVs were put
into a thermally depinned dynamic state at a higher
temperature range of $\sim$50-80 K (not shown). In the JVFB region
for 2$<$$I$$<$6~$\mu$A, JVs are represented by the combination of
the static and dynamic states [(ii)--(iv)]. The I-V
characteristics in the regime are determined by the dissipation
due to the dynamic JVs. As one moves to right in the JVFB region
the static JVs in different junctions are separately depinned and
turn into the dynamic state. Thus, the rightmost branch for
$I$$<$6~$\mu$A arises from the dynamic JVs in all 21 junctions
[(iv)].

\begin{figure}[t]
\begin{center}
\leavevmode
\includegraphics[width=0.9\columnwidth]{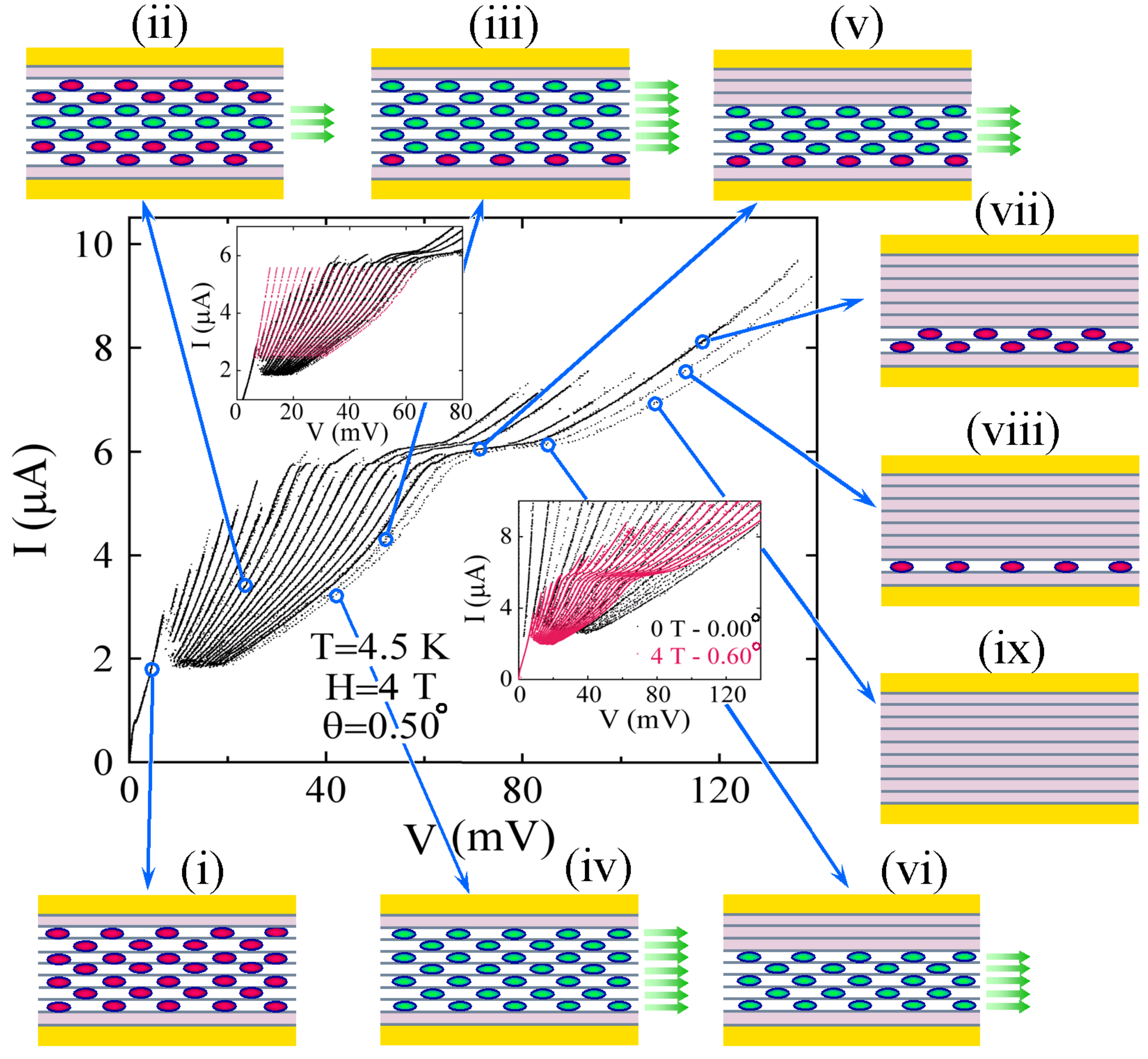}%
\caption{(colour) Schematic Josephson-vortex (JV) configurations
for the 4~T curve of Fig.\ 3(d). Some of the state configurations
are schematically illustrated, where the same legend as in Fig.\ 2
is used to denote junction vortex states. Upper inset: the voltage
spacing is uniform between neighboring JV-flow branches in a given
current, indicating that moving JVs form a triangular lattice
structure (see text for details). Lower inset: a comparison
between the branches of low-current-bias JV-flow region and the
high-current-bias extended JV region in 4~T at
$\theta$=0.6$^\circ$ (red curves), and the zero-field
quasiparticle tunneling branches (black curves), which constitutes
another evidence of dynamic triangular JV-lattice formation (see
text for details).} \label{Fig:JVScheme}
\end{center}
\end{figure}

In the bias region of $I$$\sim$6~$\mu$A, junctions with moving JVs
progressively turn into the IQT state [(v) and (vi)]. All the
junctions retaining moving JVs are resultantly in the IQT state
[(vii) to (ix)] in the extended-JV-branch region ($I$$>$7~$\mu$A).
But, the remaining junctions with static JVs, without phase
variations, are stable and thus survive even in the high-current
region. The static character of JVs deep in the extended-JV-branch
region is clearly confirmed by the lower inset of Fig.\ 4, where
the extended JV branches of 4~T at $\theta=0.6^\circ$ converge to
the zero-field IQT branches (black curves). \jydAdd{Here, only
junctions in the IQT state contribute to the tunneling voltage
while the rest of junctions containing static JVs show no
dissipation.} In the work of Ref.\ \cite{MHBaePRL07a}, no
JV-motion-induced THz emission was detected in the
extended-JV-branch region of a Bi-2212 stack, which also confirms
that only static JVs are present in the region.

Analysis of Fig.\ 4 renders the summary diagram of Fig.\ 3(f) for
different regimes of JV states in Figs.\ 3(a)--(d) as a function
of the c-axis field component of up to 350~Oe (corresponding to
$\theta$ up to 0.5$^\circ$). JJs are all in the IQT state for
$I$$>$$\sim$6~$\mu$A in $H_{(c)}$$<$$\sim$280~Oe (or
$\theta$$<$$\sim$0.4$^\circ$) [region 3]. But they turn into the
extended-JV-branch regime for $I$$>$$\sim$6~$\mu$A in
$H_{(c)}$$>$$\sim$ 280~Oe [region 4]. A transitional region exists
between the regions 2 and 4, which represents the combined state
of static and dynamic JVs, and IQT junctions.

It should be emphasized that, although the IQT branches in Fig.\
1(d) look similar to the extended JV branches in Fig.\ 4, they
correspond to totally different JV states. These IQT branches
result from the junctions retaining \textit{dynamic} JVs in
combination with IQT junctions, which is essentially same as the
state of junctions for the green curves in Fig.\ 2. Here, the
voltage difference between neighboring branches is caused by
turning the dynamic JV state (thus, a Josephson pair-tunneling
state) into the IQT state in a junction. The extended JV branches,
however, correspond to the combined state of \textit{static} JVs
and IQT junctions, where the interbranch voltage difference is due
to turning of the static JV state (also a pair-tunneling state)
into the IQT state in a junction.

In the presence of the strong inductive coupling between stacked
junctions, the static JVs are known to be stabilized in a
triangular JV lattice
\cite{BulaevskiiPRB91a,XHuPRB04a,NonomuraPRB06a,KoshelevCondMat06a}.
The structure of the moving JVs in the dynamic state, however, is
still controversial. It is widely accepted that the slowly moving
JV lattice is triangular \cite{OoiPRL02a} and becomes unstable at
a JV-flow velocity higher than the critical value, \jydAdd{which
is is close to}
the lowest mode velocity of the plasma excitation in stacked JJs
\cite{ArtemenkoPRB03a,PedersenPRB98a,HechtfischerPRB97a}. However,
other possibilities are also suggested. Those include assorted
mixture of triangular and rectangular JV lattices formed by the
resonance between the plasma excitation modes and collectively
moving JVs \cite{KleinerPRB94a,MachidaPhysicaC00a,MHBaePRB07a} or
a rectangular lattice in a certain range of external magnetic
field when the effect of boundary potential is significant
\cite{MachidaPRL06a,KoshelevPRB07a}. This study, however,
indicates that moving JVs form a triangular lattice in the entire
JVFBs region ($I<\sim6$~$\mu A$).

The upper inset of Fig.\ 4 shows a set of curves that are obtained
from the rightmost JVFB in the main panel of Fig.\ 4, on the
assumption that each junction contributes the same JV-flow voltage
for a given bias current. When these curves are overlapped on the
JVFBs of the main panel (black curves) an almost perfect scaling
of the JVFBs is seen. The equally spaced JVFBs imply that the
total JV-flow dissipation per junction, which involves both
in-plane and c-axis dissipation
\cite{KoshelevPRB00a,LatyshevPRB03a}, is same regardless of the
number of junctions with moving JVs. \jydAdd{Since the in-plane
screening current for a moving JV is time-varying it generates
finite dissipation, although it is formed in the superconducting
layers.} The in-plane dissipation in a moving \textit{rectangular
JV lattice}\jydAdd{, however,} is almost negligible except in the
top and bottom junctions
\jydAdd{because, in this JV configuration, the screening currents
induced in an extremely thin (only 0.3-nm thick) CuO$_2$ layer by
two moving JVs in neighboring junctions are cancelled with each
other.} Thus, the voltage difference between neighboring JVFBs is
expected to become larger in a more right branch for more
junctions of moving JVs. But, this is not consistent with the
observed uniform spacing between JVFBs. Moreover, a rectangular JV
lattice of vanishing in-plane dissipation should lead to the
rightmost JVFB, \jydAdd{which represents the full JV-flow
dissipation along the c axis in all the junctions}, falling on the
zero-field rightmost IQT branch. But this expectation is also in
contradiction to the observation [see the lower inset of Fig. 4].
On the other hand, the triangular lattice structure of moving JVs
with nonzero (actually maximum) in-plane \jydAdd{screening current
and} dissipation results in the dissipation that is independent of
the number of junctions with moving JVs. The finite in-plane
dissipation also explains the observed feature that the voltage of
the rightmost JVFB falls short of that of the rightmost IQT
branch. Thus, our JV-flow data provide an evidence that moving JVs
form a triangular lattice~\cite{TriangularLattice} in the entire
JVFBs region ($I<\sim6$~$\mu A$). The formation of triangular JV
lattice was previously confirmed by the magnetoresistance
oscillation \cite{OoiPRL02a} \jydAdd{for slowly flowing JVs that
are thermally depinned from the crystal edge potential} at
sufficiently high temperatures.

This picture of the triangular dynamic JV lattice, however, is not
consistent with the characteristics of the JV-motion-induced
THz-emission observed previously \cite{MHBaePRL07a}. In the study,
the frequency and the power of the emitted wave turned out to be
proportional to the total bias voltage ($V_{tot}$) over a stacked
oscillator junction and the square of the number of junctions with
moving JVs ($n$), respectively, which cannot be explained in terms
of the triangular lattice. For a moving triangular JV lattice, a
naive consideration leads to the emission frequency corresponding
to the voltage per JV-flow junction, $V_{jnc}$ (=$V_{tot}$/$n$).
Also, the emitted waves from neighboring two junctions are out of
phase, interfering destructively. In this case, the emitted power
should be oscillating with $n$ rather than monotonically
increasing in proportion to $n^2$. Thus, further examination of
the characteristics of JV-motion-induced THz emission is required
to clarify the inconsistency concerning the JV lattice structure
in the fully dynamic state.

For the static JV state the inductive coupling is weaker than the
pinning of JVs, so that depinning of JVs takes place separately in
each junction. But the inductive coupling restores its dominant
role in the dynamic JV state, for which the pinning strength is
reduced, and thereby a coherently moving JV lattice configuration
can be established.

\section{Conclusion} Stacks of Bi-2212 intrinsic Josephson junctions
sandwiched between two gold layers are employed to study the
Josephson vortex (JV) dynamics excluding the interference of the
basal part. Extreme sensitivity of the JV-flow characteristics to
the field tilt angle from the in-plane position requires the
accurate in-plane field alignment. Careful analysis of JV-flow
branches (JVFBs) obtained under different high-field strength and
slightly tilted field angles leads to the conclusion that the
JVFBs arise from combinatorial combinations of three distinct JV
states realized in different junctions and evolving separately
with magnetic field and bias current: static Josephson vortex
lattice, coherently moving lattice, and incoherent quasiparticle
tunneling state. Coherently moving JVs establish a triangular
lattice in the entire JVFBs region, as confirmed by the constant
voltage difference between JVFBs. The voltage of the rightmost
JVFB, which is smaller than that of the maximum zero-field
quasiparticle-tunneling branch in a given current, is consistent
with the formation of a triangular lattice configuration in the
dynamic state. Here, the difference in the voltage is caused by
the in-plane dissipation by the JVs in motion. Information on the
JV dynamic states and interaction between JVs and PVs can be
conveniently utilized to control the JV motion in stacked
Josephson junctions.

\acknowledgements HJL acknowledges the valuable discussion with
M.~Machida. We appreciate U. Welp for critical reading and
valuable suggestions. This work was supported by Acceleration
Research Grant (No.\ R17-2008-007-01001-0) administered by Korea
Science and Engineering Foundation. Work in Argonne was supported
by U.S. DoE Office of Science under Contract No.
DE-AC02-06CH11357.

\bibliography{JVFBs}

\begin{thebibliography}{40}
\bibitem[{QTB()}]{QTBs}
\bibinfo{note}{Since natural Josephson junctions in Bi-2212 single crystals
  have underdamped characteristics each junction presents a quasiparticle
  tunneling branch in the hysteretic I-V curves.}
\bibitem[{Pin()}]{PinningOrigin}
\bibinfo{note}{Inhomogeneous critical current density due to crystal defects or
  bounday potential of the sample are possible origin of JV pinning.}
\bibitem[{Tri()}]{TriangularLattice}
\bibinfo{note}{Finite in-plane dissipation is also possible for any sheared JV
  lattice structures. But it has been theoretically predicted that moving JVs
  are stabilized in a triangular lattice~\cite{ArtemenkoPRB03a}.}
\end{thebibliography}

\end{document}